\documentclass[twocolumn,twocolappendix]{aastex6}
\usepackage{times,epsfig,amssymb,amsmath}
\usepackage{graphicx}
\usepackage{color}
\usepackage{xparse}
\usepackage{tablefootnote}
\usepackage{ulem}
\fullcollaborationName{The Friends of AASTeX Collaboration}

\newcommand{\Planck}{{\em Planck}}
\newcommand{\XMM}{XMM-{\em Newton}}
\newcommand{\OmegaM}{\Omega_{\rm m}}

\newcommand{\Mpl}{M^\text{Pl}_{200}}
\newcommand{\mMpl}{M^\text{Pl}_{500}}
\newcommand{\Mt}{M_\text{200}}
\newcommand{\sigv}{\sigma_\text{1D}}
\newcommand{\lsigv}{s_\text{v}}
\newcommand{\lMpl}{s_\text{Pl}}

\newcommand{\av}{a_\text{v}}
\newcommand{\alphav}{\alpha_\text{v}}

\newcommand{\Ag}{A_\text{g}}
\newcommand{\ag}{{a_\text{g}}}

\newcommand{\alphag}{{\alpha_\text{g}}}
\newcommand{\Ad}{A_\text{d}}
\newcommand{\ad}{{a_\text{d}}}

\newcommand{\alphad}{\alpha_\text{d}}
\newcommand{\Msol}{M_\odot}
\newcommand{\disv}{\tilde{\sigma}_{\lsigv}}
\newcommand{\disMpl}{\tilde{\sigma}_{\lMpl}}
\newcommand{\Tdisv}{\Sigma_{\lsigv}}
\newcommand{\TdisMpl}{\Sigma_{\lMpl}}
\newcommand{\Mdisv}{\sigma_{\lsigv}}
\newcommand{\MdisMpl}{\sigma_{\lMpl}}
\newcommand{\mlsigv}{\bar{\lsigv}}
\newcommand{\mlMpl}{\bar{\lMpl}}
\newcommand{\corr}{\tilde{r}}
\newcommand{\feb}{f_\text{EB}}
\newcommand{\fr}{f_\text{corr}}
\newcommand{\bvel}{b_\text{v}}

\begin{document}


\title{Calibrating the {\em PLANCK} Cluster Mass Scale with Cluster Velocity Dispersions}


\author{Stefania Amodeo\altaffilmark{1},
Simona Mei\altaffilmark{1,2,3,4,8}, 
Spencer A. Stanford\altaffilmark{5},
James G. Bartlett\altaffilmark{3,6},
Jean-Baptiste Melin\altaffilmark{7},
Charles R. Lawrence\altaffilmark{3},
Ranga-Ram Chary\altaffilmark{8},
Hyunjin Shim\altaffilmark{9,8},
Francine Marleau\altaffilmark{10,8},
Daniel Stern\altaffilmark{3}
}
\altaffiltext{1}{LERMA, Observatoire de Paris, PSL Research University, CNRS, Sorbonne Universit\'es, UPMC Univ. Paris 06, F-75014 Paris, France} 
\altaffiltext{2}{Universit\'{e} Paris Denis Diderot, Universit\'e Paris Sorbonne Cit\'e, 75205 Paris Cedex
13, France}
\altaffiltext{3}{Jet Propulsion Laboratory, California Institute of Technology, 4800 Oak Grove Drive, Pasadena, California, USA}
\altaffiltext{4}{Cahill Center for Astronomy \& Astrophysics, California Institute of Technology, Pasadena, CA 91125, USA}
\altaffiltext{5}{Department of Physics, University of California Davis, One Shields Avenue, Davis, CA 95616, USA ; Institute of Geophysics and Planetary Physics, Lawrence Livermore National Laboratory, Livermore, CA 94550, USA}
\altaffiltext{6}{APC, AstroParticule et Cosmologie, Universit\'e Paris Diderot, CNRS/IN2P3, CEA/lrfu, Observatoire de Paris, Sorbonne Paris Cit\'e, 10 rue Alice Domon et L\'eonie Duquet, 75205, Paris Cedex 13, France}
\altaffiltext{7} {DRF/Irfu/SPP, CEA-Saclay, 91191, Gif-sur-Yvette Cedex, France}
\altaffiltext{8} {Infrared Processing and Analysis Center, California Institute of Technology, Pasadena, CA 91125, USA}
\altaffiltext{9} {Department of Earth Science Education, Kyungpook National University, Republic of Korea}
\altaffiltext{10}{Institute of Astro and Particle Physics, University of Innsbruck, 6020, Innsbruck, Austria}







\begin{abstract}
We measure the \Planck\ cluster mass bias using dynamical mass measurements based on velocity dispersions of a subsample of 17 \Planck-detected clusters.  The velocity dispersions were calculated using redshifts determined from spectra obtained at Gemini observatory with the GMOS multi-object spectrograph.  We correct our estimates for effects due to finite aperture, Eddington bias and correlated scatter between velocity dispersion and the \Planck\ mass proxy.  The result for the mass bias parameter, $(1-b)$, depends on the value of the galaxy velocity bias $\bvel$ adopted from simulations: $(1-b)=(0.51\pm0.09) \bvel^3$.
Using a velocity bias of $\bvel=1.08$ from Munari et al., we obtain $(1-b)=0.64\pm 0.11$, i.e, an error of 17\% on the mass bias measurement with 17 clusters. This mass bias value is consistent with most previous weak lensing determinations. It lies within $1\sigma$ of the value needed to reconcile the \Planck\ cluster counts with the \Planck\ primary CMB constraints. We emphasize that uncertainty in the velocity bias severely hampers precision measurements of the mass bias using velocity dispersions. On the other hand, when we fix the \Planck\  mass bias using the constraints from Penna-Lima et al., based on weak lensing measurements, we obtain a positive velocity bias $\bvel \gtrsim 0.9$ at $3\sigma$.
\end{abstract}




\section{Introduction} \label{sec:intro}
Galaxy clusters are fundamental tools for tracing the evolution of cosmic structures and constraining cosmological parameters.  Their number density  at a given epoch is strongly dependent on the amplitude of density fluctuations, $\sigma_8$ (the standard deviation within a comoving sphere of radius 8$h^{-1}$\,Mpc), and the matter density of the Universe, $\OmegaM$ \citep[see, e.g., the review by][]{allen+11}.  A key quantity for using galaxy clusters as cosmological probes is their mass.  Unfortunately, mass is not directly observable, but it can be estimated through several, independent methods based on different physical properties, each affected by its own set of specific systematic effects.  Methods are based on the analysis of the thermal emission of the intracluster medium (ICM), observed either in the X-rays or through the Sunyaev-Zeldovich (SZ) effect \citep{sz1970}, the dynamics of member galaxies and gravitational lensing.  Comparison of mass estimates by different techniques is a critical check on the reliability of each method under different conditions, and also a test of the cosmological scenario.  

The SZ effect originates from the transfer of energy from the heated electrons in the ICM to the photons of the cosmic microwave background (CMB) via inverse Compton scattering \citep[see review by][]{carlstrom2002}.  This scattering generates a distortion of the blackbody spectrum of the CMB that appears as a decrease in intensity at frequencies below 218\,GHz and an increase at higher frequencies.  The amplitude of the effect is quantified by the Compton parameter integrated along the line-of-sight, $y \propto T_{\rm e} n_{\rm e}$, where $T_{\rm e}$ and $n_{\rm e}$ are the electron temperature and density, respectively; or equivalently by its solid-angle integral, $Y = \int y \, d\Omega$.  Unlike optical or X-ray emission, the surface brightness of the SZ effect (relative to the mean CMB brightness) is independent of distance.  Dedicated SZ cluster surveys can therefore efficiently find clusters out to high redshifts. Moreover, since the SZ signal is proportional to the thermal energy of the ICM, it can be used to estimate total cluster mass, and numerical simulations \citep[e.g.][]{kravtsov+06} show that the integrated Compton signal, $Y$, tightly correlates with the mass.

Recent millimeter-wave surveys are providing large samples of SZ-detected clusters and applying them in cosmological analysis: the South Pole Telescope \citep[SPT;][]{bleem+15, deHaan+16}, the Atacama Cosmology Telescope \citep[ACT;][]{marriage+11, hasselfield+13} and the \Planck\ satellite \citep{planck_xxxii}.  \Planck\ produced two all-sky SZ cluster catalogs, the PSZ1 with 1227 detections based on 15.5 months of data, and the PSZ2 with 1653 detections from the full mission dataset of 29 months \citep{Planck2014, Planck2015}.  Using subsamples of confirmed clusters at higher detection significance, \Planck\  constrained cosmological parameters from the cluster counts \citep{PlanckSZCosmo2014, planck_xxiv}, noting tension with the values of $\sigma_8$ and $\OmegaM$ favored by the primary CMB anisotropies.  

The largest source of uncertainty in cosmological inference from the cluster counts is the SZ-signal-halo mass relation. Higher angular resolution SZ observations show that the \Planck\ determination of the SZ signal is robust \citep{rg+15,sayers+15}.  \Planck\ calibrates the  relation with mass proxies from \XMM\ X-ray observations \citep{UPP2010}, proxies that are in turn calibrated assuming hydrostatic equilibrium of the ICM \cite[see the Appendix of][]{PlanckSZCosmo2014}.  This assumption, however, neglects possible contributions from bulk motions and non-thermal sources to the pressure support of the ICM.  Analyses of mock data from simulations indicate that these can cause a 10-25\% underestimate of cluster total mass \citep[e.g.,][]{nagai+07, PV2008, meneghetti+10}.  Other effects, such as instrument calibration or temperature inhomogeneities in the gas \citep{rasia+06, rasia+14}, can additionally bias hydrostatic mass measurements.  It is common to lump all possible astrophysical and observational biases into the {\it mass bias parameter}, $(1-b)$, defined in Section \ref{sec:results}.   Simulations and comparison of different X-ray analyses indicate the range $b=0-40\%$, with a baseline value of 20\% \citep{mazzotta+04, nagai+07, PV2008, lau+09, kay+12, rasia+12,rozo+14a,rozo+14b,rozo+14c}.  To reconcile the \Planck\ cluster constraints with those of the primary CMB requires a mass bias of $(1-b) = 0.58\pm 0.04$ \citep{planck_xxiv}.

Weak gravitational lensing (WL) provides an alternate method of measuring cluster mass \citep[e.g.,][]{hj2008}.  Bending of light by the cluster gravitational field distorts the images of background galaxies, elongating them tangentially around the cluster.  Statistical analysis of such distortions gives a direct estimate of the density profile of the cluster and its total mass. Gravitational lensing is particularly efficient in estimating cluster mass because it is sensitive to the total mass, independently of cluster composition or dynamical state. However, since WL measures the projected mass, cluster triaxiality and the presence of substructures along the line-of-sight introduce significant noise; nevertheless, the noise can be reduced by stacking the WL signal from a large number of clusters to yield an un-biased estimate of the sample mass \citep[][]{sheldon+04,johnston+07,ck2009, meneghetti+10, bk2011}.  

Several recent WL calibrations of the \Planck\ cluster scale have found results in the range $0<b<30$\%, at the 10\% precision level \citep{WtG,CCCP,simet+15,smith+16}.  \citet{mb15} propose a new technique to measure cluster masses through lensing of CMB temperature anisotropies, and \citet{Planck2015}, \citet{baxter+15} for SPT and \citet{madhavacheril+15} for ACT all report first detections of this effect that holds great promise for the future.  \citet{battaglia+15} have pointed out the potential impact of Eddington bias -- the steep mass function scattering more low than high mass objects into an SZ signal bin -- on these mass calibrations.  Using a complete Bayesian analysis to account for this and other effects, \citet[][]{penna-lima+2016} obtained a value of $b\sim 25$\%, consistent with previous measurements.  All of this illustrates the importance of cluster mass measurements and the need for independent determinations and increasing precision.  

An additional, widely used method to constrain cluster mass takes the velocity dispersion of member galaxies as a measure of the gravitational potential of the dark matter halo, assumed to be in virial equilibrium. The scaling relation between velocity dispersion and mass has been well established by cosmological N-body and hydrodynamical simulations \citep[e.g.,][]{evrard+08, munari+13}, which confirm the trend $\sigma \propto M^{1/3}$ expected from the virial relation for a broad range of masses, redshift and cosmological models.  Cluster member galaxies may not, however, share the same velocity dispersion as the bulk of the dark matter, because they are hosted by subhalos whose dynamical state may differ.  This introduces the concept of velocity bias \citep[e.g.,][]{carlberg1994,colin+00} that mass estimates must account for.  Recently, \citet{sifon+16} presented dynamical mass estimates based on galaxy velocity dispersions for a sample of 44 clusters observed with ACT.  Their sample spans a redshift range $0.24<z<1.06$, with an average of 55 spectroscopic members per cluster. Comparing dynamical and SZ mass estimates, they find a mass bias of $(1-b)=1.10\pm0.13$ (i.e., $b=-10$\%).

In the present work, we study the relation between velocity dispersion and the SZ \Planck\ mass for a sample of 17 \Planck\ clusters observed at the Gemini Observatory to estimate the mass bias parameter.  All but one are in the PSZ2.  In Section \ref{sec:dataset} we describe the observations and the sample, and then present our results in Section \ref{sec:results}.  We discuss the resulting mass bias measurement and compare our results to previous measurements in Section \ref{sec:discussion}; we also turn the analysis around to constrain the velocity bias by adopting a constraint on the mass bias from WL observations.   Section \ref{sec:conclusions} concludes.  Throughout, we adopt the \Planck\ base $\Lambda$CDM model \citep{Planck2015}: a flat universe with $\OmegaM=0.307$ and $H_0=67.74$\,km\,s$^{-1}$\,Mpc$^{-1}$ ($h\equiv H_0/(100$\,km\,s$^{-1}$\,Mpc$^{-1}$).  Mass measurements are quoted at a radius $R_\Delta$, within which the cluster density is $\Delta$ times the critical density of the universe at the cluster's  redshift, where $\Delta = \left\{200, 500\right\}$.  All quoted uncertainties are 68.3\% (1$\sigma$) confidence level, unless otherwise stated.
\section{The dataset}
\label{sec:dataset}

\subsection{Gemini/GMOS spectroscopy}
\label{sec:Gemini}
The goal of our program was to obtain an independent statistical calibration of the \Planck\ SZ mass estimator.  We chose \Planck\ SZ-selected clusters that were detected with a signal-to-noise of 4.5~$\sigma$ or larger, distributed in the North and in the South, and with a broad range in mass. 
We obtained pre-imaging and optical spectroscopy with GMOS-N and GMOS-S at the Gemini-North and Gemini-South Telescopes (Programs GN-2011A-Q-119, GN-2011B-Q-41, and GS-2012A-Q-77; P.I. J.G. Bartlett), respectively, of 19 galaxy clusters, spanning a range in \Planck\ SZ masses of $2\times 10^{14} M_{\sun} \lesssim M_{500,\text{SZ}} \lesssim 10^{15} M_{\sun}$ (a more detailed discussion of these observations will follow in a companion paper). We were able to obtain velocity dispersion measurements for 17 clusters, which constitute our sample in this paper. All but one (CL G183.33-36.69) are in the PSZ2 catalog.

The Northern sample was selected in the SDSS \citep[Sloan Digital Sky Survey;][]{york+00} area. We used the SDSS public releases and GMOS-N pre-imaging in the {\it r}-band for 150~s to detect red galaxy over-densities at the \Planck\ detection, and, when unknown, estimate the approximate redshift using their red sequence. For PSZ2 G139.62+24.18 and PSZ2 G157.43+30.34, we used imaging obtained with the Palomar telescope  (PI: C. Lawrence).  
For the Southern sample, we obtained GMOS-S imaging in the {\it g} and {\it i}-bands for 200~s and 90~s, respectively. 
Red galaxy over-densities and cluster members were selected by their colors, using \citet{BC2003} stellar population models and \citet{mei+09} empirical red sequence measurements.  In Table~\ref{sample}, we list our sample properties and the spectroscopy observing times. 

The GMOS spectra were reduced using the tasks in the IRAF Gemini GMOS package and standard longslit techniques.  After co-adding the reduced exposures, one-dimensional spectra for the objects in each slitlet were extracted and initially inspected visually to identify optical features such as the 4000 \AA ~break, G-band, Ca H$+$K absorption lines, and, rarely, [O~II]$\lambda$3727.   More precise redshifts were determined by running the IRAF xcsao task on these spectra. 
We calculate the cluster velocity dispersions using the ROSTAT software \citep{beers+90} with both the Gaussian and biweight methods, which are appropriate to our clusters where there are typically 10 -- 20  confirmed members. We retain as cluster members galaxies within 3$\sigma$ of the average cluster redshift. 
From the original sample of 19 clusters, we have excluded two, which have complex non-Gaussian velocity distribution profiles.  In a companion paper (Amodeo et al. 2017b, in prep), we show the velocity histograms of all observed clusters and publish catalogs of spectroscopic redshift measurements.

An important assumption that we make for this analysis is that our cluster sample is a representative, random sub-sample of the \Planck\ SZ selected catalogue.  In this case there are no corrections for selection effects, such as Malmquist bias, because we determine the mean scaling for velocity dispersion {\em given} the SZ mass proxy.

\begin{table*}
\begin{center}
\caption{The cluster sample used in this paper. We list the PSZ2 cluster ID, when available. When it is not available, we use the prefix 'CL' followed by a notation in Galactic coordinates similar to that used in the PSZ2 paper.  \label{sample}}
\vspace{0.25cm}
\resizebox{!}{5.5cm}{
\begin{tabular}{llccccccccccccccc}
\tableline \tableline\\
Name & R.A.&decl.&Im. filter&$t_{\text{exp}}$&$N_{\text{mask}}$&Run\\
&(deg)&(deg)&&(s)&  &\\
\tableline \tableline \\
PSZ2 G033.83-46.57 &326.3015&-18.7159&\it g,i&1800&2&GS-2012A-Q-77\\
PSZ2 G053.44-36.25 &323.8006&-1.0493&\it r&1800&1&GN-2011A-Q-119,GN-2011B-Q-41\\
PSZ2 G056.93-55.08 &340.8359&-9.5890&\it r&1800&2&GN-2011A-Q-119,GN-2011B-Q-41\\
PSZ2 G081.00-50.93 &347.9013&3.6439&\it r&1800&&GN-2011A-Q-119,GN-2011B-Q-41\\
PSZ2 G083.29-31.03 &337.1406&20.6211&\it r&1800&&GN-2011A-Q-119,GN-2011B-Q-41\\
PSZ2 G108.71-47.75 &3.0715 &14.0191&\it r&1800&2&GN-2011A-Q-119,GN-2011B-Q-41\\
PSZ2 G139.62+24.18 &95.4529&74.7014&\it r&900&2&GN-2011A-Q-119,GN-2011B-Q-41\\
&&&\it g,i,r,J,K&&&Palomar Hale Telescope\\
PSZ2 G157.43+30.34 &117.2243&59.6974&\it r&3600&2&GN-2011A-Q-119,GN-2011B-Q-41\\
&&&\it g,i,r,J,K&&&Palomar Hale Telescope\\
CL G183.33-36.69 &57.2461&4.5872&\it r&1800&2&GN-2011A-Q-119,GN-2011B-Q-41\\
&&&\it g, J, K&&&Palomar Hale Telescope\\
PSZ2 G186.99+38.65 &132.5314&36.0717&\it r&1800&2&GN-2011A-Q-119,GN-2011B-Q-41\\
PSZ2 G216.62+47.00 &147.4658&17.1196&\it r&1800&2&GN-2011A-Q-119,GN-2011B-Q-41\\
PSZ2 G235.56+23.29 &134.0251&-7.7207&\it g,i&900&2&GS-2012A-Q-77\\
PSZ2 G250.04+24.14&143.0626&-17.6481&\it g,i&1800&&GS-2012A-Q-77\\
PSZ2 G251.13-78.15 &24.0779&-34.0014&\it g,i&900&2&GS-2012A-Q-77\\
PSZ2 G272.85+48.79 &173.2938&-9.4812&\it g,i&900&2&GS-2012A-Q-77\\
PSZ2 G329.48-22.67 &278.2527&-65.5555&\it g,i&900&2&GS-2012A-Q-77\\
PSZ2 G348.43-25.50 &291.2293&-49.4483&\it g,i&900&2&GS-2012A-Q-77\\
\tableline \tableline\\
\end{tabular}}
\end{center}
\end{table*}

\begin{table*}
\begin{center}
\caption{Columns from left to right list the cluster ID, our measured average redshift, the number of confirmed member galaxies, the maximum radius probed by GMOS, $R_{\text{max}}$, $R_\text{200}$, our measured velocity dispersion, $\sigma(<R_{\text{max}})$, the velocity dispersion estimated within $R_\text{200}$, $\sigma_{200}$, the reference PSZ2 $\mMpl\ $ and the $\Mpl\ $ derived in this work based on SZ. \label{results}}
\vspace{0.25cm}
\resizebox{!}{5cm}{
\begin{tabular}{llccccccccccccccc}
\tableline \tableline\\
Name &$z$&$N_{\rm gal}$&$R_{\text{max}}$&$R_\text{200}$&$\sigma_{1D}(<R_{\text{max}})$&$\sigma_{200}$&$\Mpl\ $&$\mMpl\ $ \\
& &&($R_\text{200}$)&(Mpc)&(km~$\rm s^{-1}$)&(km~$\rm s^{-1}$)&($10^{14} M_\odot $)&($10^{14} M_\odot $)\\
\tableline \tableline \\
PSZ2 G033.83-46.57 &0.439&10&$0.58$&$1.66\pm0.08$&$985\substack{+451\\-277}$&$953\substack{+454\\- 282}$&$7.8\pm1.1$&$5.4\substack{+0.7\\-0.8}$\\
PSZ2 G053.44-36.25 &0.331&20&$0.42$&$1.93\pm 0.06$&$1011\substack{+242\\-131}$&$956\substack{+260\\-161}$&$10.9\pm1.0$&$7.5\substack{+0.5\\-0.6}$\\
PSZ2 G056.93-55.08 &0.443&46&$0.49$&$2.00\pm 0.05$&$1356\substack{+192\\-127}$&$1290\substack{+218\\-164}$&$13.8\pm1.1$&$9.4\pm0.5$\\
PSZ2 G081.00-50.93 &0.303&15&$0.41$&$1.88\pm 0.06$&$1292\substack{+360\\-185}$&$1220\substack{+381\\-223}$&$9.8\pm0.9$&$6.7\pm0.5$\\
PSZ2 G083.29-31.03 &0.412&20&$0.49$&$1.89\pm 0.06$&$1434\substack{+574\\-320}$&$1365\substack{+584\\-338}$&$11.3\pm1.0$&$7.8\substack{+0.5\\-0.6}$\\
PSZ2 G108.71-47.75 &0.390&10&$0.55$&$1.65\pm 0.08$&$900\substack{+458\\-190}$&$865\substack{+461\\-198}$&$7.3\pm1.1$&$5.1\substack{+0.7\\- 0.8}$\\
PSZ2 G139.62+24.18 &0.268&20&$0.36$&$1.96\pm 0.06$&$1120\substack{+366\\-238}$&$1052\substack{+390\\-273}$&$10.6\pm0.9$&$7.3\pm0.5$\\
PSZ2 G157.43+30.34 &0.402&28&$0.47$&$1.94\pm 0.05$&$1244\substack{+192\\-109}$&$1182\substack{+216\\-148}$&$12.1\pm1.0$&$8.2\pm0.6$\\
CL G183.33-36.69 &0.163&11&$0.35$&$1.38\pm 0.17$&$897\substack{+437\\-275}$&$842\substack{+451\\-297}$&$3.3\pm1.2$&$2.3\substack{+0.7\\-0.9}$\\
PSZ2 G186.99+38.65 &0.377&41&$0.49$&$1.81\pm 0.06$&$1506\substack{+164\\-120}$&$1432\substack{+200\\-166}$&$9.5\pm1.0$&$6.6\substack{+0.6\\-0.7}$\\
PSZ2 G216.62+47.00 &0.385&37&$0.45$&$1.97\pm 0.05$&$1546\substack{+174\\-132}$&$1466\substack{+218\\-186}$&$12.3\pm1.0$&$8.4\substack{+0.5\\-0.6}$\\
PSZ2 G235.56+23.29 &0.374&23&$0.51$&$1.73\pm 0.08$&$1644\substack{+285\\-192}$&$1568\substack{+ 308\\-224}$&$8.2\pm1.2$&$5.7\substack{+0.7\\-0.8}$\\
PSZ2 G250.04+24.14 &0.411&29&$0.53$&$1.75\pm 0.07$&$1065\substack{+447\\-285}$&$1020\substack{+452\\-293}$&$8.9\pm1.0$&$6.2\pm0.6$\\
PSZ2 G251.13-78.15 &0.304&9&$0.48$&$1.59\pm 0.08$&$ 801\substack{+852\\-493}$&$762\substack{+854\\-497}$&$5.9\pm0.9$&$4.1\pm0.6$\\
PSZ2 G272.85+48.79 &0.420&10&$0.57$&$1.65\pm 0.08$&$1462\substack{+389\\-216}$&$1411\substack{+397\\-231}$&$7.6\pm1.1$&$5.3\substack{+0.7\\-0.8}$\\
PSZ2 G329.48-22.67 &0.249&11&$0.38$&$1.73\pm 0.07$&$ 835\substack{+179\\-119}$&$786\substack{+200\\-149}$&$7.2\pm	0.9$&$5.0\substack{+0.5\\-0.6}$\\
PSZ2 G348.43-25.50 &0.265&20&$0.37$&$1.84\pm 0.06$&$1065\substack{+411\\-198}$&$1003\substack{+427\\-230}$&$8.7\pm0.9$&$6.0\pm0.6$\\
\tableline \tableline\\
\end{tabular}}
\end{center}
\end{table*}

\subsection{\Planck\ Mass Proxy}
\label{sec:Mpl}
The \Planck\ SZ mass proxy is based on a combination of \Planck\ data and an X-ray scaling relation established with \XMM.  It has been used in the last two \Planck\ cluster catalog papers \citep{Planck2014,Planck2015}.  Here we give a brief summary and refer the reader to Sect. 7.2.2 of \cite{Planck2014} for more details.

With respect to the PSZ2, in this paper we derive new cluster mass estimates, taking into account the cluster centers from our Gemini/Palomar optical follow-up.
For each cluster, we measure the SZ flux, $Y_{500}$, inside a sphere of radius $R_{500}$ using the Multifrequency Matched Filter \citep[MMF3,][]{MMF2006}.  The filter combines the six highest frequency bands (100-857\,GHz) weighted to optimally extract a signal with the known SZ spectral shape and with an assumed spatial profile.  For the latter, we adopt the so-called universal pressure profile from \citet{UPP2010}.  We center the filter on the optical position and vary its angular extent $\theta_{500}$ over the range [0.9 - 35] arcmin to map out the signal-to-noise surface over the flux-size ($Y_{500}-\theta_{500}$) plane.  In the \Planck\ data there is a degeneracy between the measured flux and cluster size defined by this procedure, which we break using an X-ray determined scaling relation as a prior constraint \citep[i.e., an independent $Y-\theta$ relation obtained from the combination of Eq. 7 and 9 of][]{PlanckSZCosmo2014}.  The intersection of this prior with the \Planck\ degeneracy contours yields a tighter constraint on the flux $Y_{500}$, which we then convert to halo mass, $\mMpl$, using Eq.~(7) of \cite{PlanckSZCosmo2014}.  It is important to note that the mass proxy is therefore calibrated on the \XMM\ scaling relation.  These masses are reported in Table~\ref{results}.
In order to compare our mass measurements to other independent estimates, we rescale the \Planck\ masses to $\Mpl\ $ using the mass-concentration relation of \citet{dm14}. The rescaling procedure is described in Appendix~\ref{sec:conversion} and the resulting values of $\Mpl\ $ are listed in Table~\ref{results}.

\subsection{Correcting velocity dispersions for GMOS finite aperture}
\label{sec:CorrVel}
The GMOS spectrographs provide imaging and spectroscopy over a 5.5x5.5 $\text{arcmin}^2$ field of view, allowing measurements for only the central part of clusters. The radial coverage provided for each cluster at a given redshift, calculated for the \Planck\ 2015 cosmology, is quoted in Table \ref{results} as $R_\text{max}$, in units of $R_\text{200}$, along with $R_\text{200}$. We typically sample out to about half $R_{200}$, with $R_\text{max}$ ranging over $[0.35-0.58] R_\text{200}$. However, we need to estimate the velocity dispersion within $R_\text{200}$,  $\sigma_\text{200}\equiv \sigma(<R_\text{200})$, to compare to the $\sigma$--M relation from simulations (see next section). \citet{sifon+16} determine the radial profile of the velocity dispersion using mock observations of subhalos in the Multidark simulation \citep{prada+12}, as described in Section 3.2 of their paper. We interpolate the correction factors presented in their Table 3 to our values of $R_\text{max}/R_\text{200}$ to translate our velocity dispersion measurements, $\sigv(<R_\text{max})$, to $\sigma_\text{200}$. The velocity dispersions thusly estimated are listed in Table \ref{results}, where the uncertainties account for our measurement errors and the scatter in the velocity dispersion profile found by \citet{sifon+16}. The mean corrections are of order 5\%, while the uncertainty increases up to 32\%. Figure~\ref{fig:mass_sig} plots the velocity dispersions  $\sigma_{200}$ versus $\Mpl\ $.

\section{Analysis: the mass bias}
\label{sec:results}
\subsection{The mass bias and the velocity bias}
Our goal is to find the \Planck\ cluster mass scale using velocity dispersion as an independent mass proxy calibrated on numerical simulations.  We define the mass bias factor, $(1-b)$, in terms of the ratio between the \Planck-determined mass, $\Mpl$, and true cluster mass, $\Mt$ \citep{Planck2015,WtG,CCCP}. We assume that it is a constant and independent of over-density. In fact, while the mass bias may depend on mass and other cluster properties, our small sample only permits us to constrain a characteristic value averaged over the sample.   For $M_{200}$  the mass bias is defined as:
\begin{equation}\label{eq:massbias}
\Mpl = (1-b) \Mt \,.
\end{equation}

Complete virialization predicts a power-law relation between velocity dispersion, $\sigma_\text{200}$, and mass, $\Mt$. 
Following the approach used in simulations, we work with the logarithm of these quantities, $\lsigv = \ln(\sigma_\text{200}/\text{km\,s}^{-1})$, $\mu = \ln(E(z)\Mt/10^{15}\,\Msol)$, where $h(z)\equiv H(z)/(100$\,km\,s$^{-1}$\,Mpc$^{-1})=hE(z)$ is the dimensionless Hubble parameter at redshift $z$, and we consider the log-linear relation
\begin{equation}\label{eq:dmsig}
\langle \lsigv | \mu \rangle  =  a_\text{d} + \alphad\mu \,.
\end{equation}
The so-called self-similar slope expected from purely gravitational effects is $\alphad=1/3$.
The angle brackets indicate that this is the mean value of $\lsigv$ given $\mu$.  From a suite of  simulations, \citet{evrard+08} determined a precise relation between the dark matter velocity dispersion and halo mass consistent with this expectation. They find a normalization $\ad = \ln\left(1082.9\pm4.0 \right)+\alphad\ln h$; in the following, we will also refer to $\Ad\equiv e^\ad$.
The result is insensitive to cosmology and to non-radiative baryonic effects, 
and the relation is very tight with only 4\% scatter at fixed mass.

Galaxies, however, may have a different velocity dispersion than their dark matter host because they inhabit special locations within the cluster, e.g., subhalos.  This leads to the concept of velocity bias, in which the scaling of galaxy velocity dispersion with host halo mass will in general be fit by a relation of the form of Eq.~(\ref{eq:dmsig}), but with different parameters, $\Ag\equiv e^\ag$ and $\alphag$.  Simulations typically find the exponent $\alphag$ to be consistent with the self-similar value of $1/3$, so we can quantify any velocity bias in terms of the normalization, $\Ag$.  We do so by introducing the velocity bias parameter, $\bvel\equiv \Ag/\Ad$.

Different simulation-based or empirical analyses find discordant behavior for the velocity bias, leaving even the sense of the effect (i.e., $\bvel>1$ or $\bvel<1$) in debate.  

Using hydrodynamical simulations with star formation, gas cooling and heating by supernova explosions and AGN feedback, \citet{munari+13} find that subhalos and galaxies have a slightly {\em higher} velocity dispersion than the dark matter, i.e., a {\em positive} velocity bias with $\bvel>1$.  For galaxies in their AGN-feedback model, for example, they find $\tilde{\Ag}=1177$,  corresponding to $\bvel=1.08$. 

From combined N-body and hydrodynamical simulations, \citet{wu+13} find that velocity bias depends on the tracer population; in particular, that subhalos in pure N-body simulations tend to have large positive bias compared to galaxies identified in the hydrodynamical simulations, perhaps because over-merging in the former case removes slower, low mass dark matter halos from the tracer population.  Consistent with this picture where smaller objects are more efficiently destroyed, all tracers in their simulations show increasingly positive velocity bias with decreasing subhalo mass or galaxy luminosity, independent of redshift. The brightest cluster galaxies tend to underestimate, and faint galaxies slightly overestimate, the dark matter halo velocity dispersion, with the velocity bias ranging from $\sim$0.9 for the five brightest cluster galaxies to an asymptotic value of $\bvel=1.07$ when including the 100 brightest galaxies (see Figure 1 in their paper). For samples of more than $\sim 50$ galaxies, their result converges to the value of \citet{munari+13} ($\bvel=1.08$). The 10-20 brightest galaxies, similar to our observational sample, represent a nearly unbiased measurement of the halo velocity dispersion, i.e., $\bvel=1$. 

On the other hand, \citet{guo+15} observe the opposite trend with luminosity when measuring the velocity bias of galaxies in the Sloan Digital Sky Survey (SDSS) Data Release 7 (see their Figure 9). They find $\bvel\simeq1.1$ for the brightest galaxies, falling to 0.85 for faint galaxies. It is worth noting that this analysis is based on modeling of the projected and redshift-space two-point correlation functions, and it is probably not very sensitive to velocity bias in the most massive halos, such as we have in the \Planck\ sample. \citet{farahi+16} use the velocity bias from the bright subsample of \citet{guo+15} ($\bvel=1.05\pm0.08 $) to estimate the mass of redMaPPer clusters with stacked galaxy velocity dispersions. Their derived mass scale is consistent with estimates based on weak lensing observations reported by \citet{simet+16}. The \citet{guo+15} observational result is also consistent with the value $\bvel = 1.08$ from the N-body hydrodynamical simulations of \citet{munari+13}.  In an another study, \citet{caldwell+16} find a negative velocity bias, $\bvel=0.896$, for galaxies in their simulations when they adjust feedback efficiencies to reproduce the present-day stellar mass function and the hot gas fraction of clusters and groups.

These different studies do not yet present a clear picture of the magnitude of cluster member velocity bias, and this quantity remains the primary factor limiting interpretation of dynamical cluster mass measurements at present. We use the Munari et al. value of the velocity bias, $\bvel=1.08$, as our baseline in the following. The uncertainty on Munari et al.'s velocity bias is $\sim 0.6\%$.

\subsection{Measurement of the mass bias}

\begin{figure*}
\figurenum{1}
\plotone{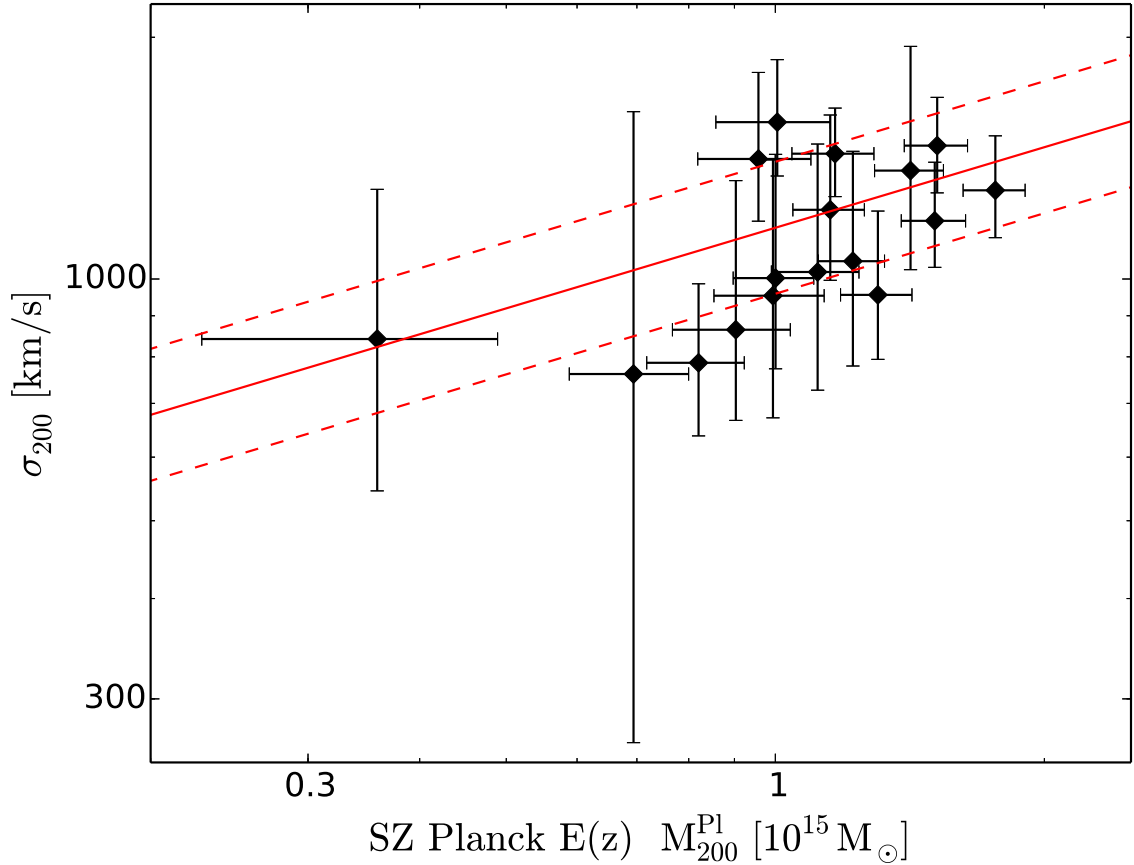}
\caption{Relation between the \Planck\ SZ mass proxy and velocity dispersion for our sample of 17 galaxy clusters observed with Gemini (diamonds).  The velocity dispersions and the \Planck\ masses have been converted to $\sigma_\text{200}$ and $\Mpl$, respectively, with corresponding uncertainties, following the procedure described in the text.  The solid red line shows the best fit to the functional form of Eq.~(\ref{eq:dmsig}) in log-space, where the slope is set to 1/3, with the dashed lines delineating the dispersion of the data about the best-fit line.}
\label{fig:mass_sig}
\end{figure*}

As detailed in Appendix~\ref{sec:appendix}, our model of constant mass bias, $(1-b)$, predicts a log-linear scaling relation of the form Eq.~(\ref{eq:dmsig}) between the observed velocity dispersion and the \Planck \ mass proxy.  We therefore construct an estimator for $(1-b)$ by fitting for the normalization, $a$, and exponent, $\alpha$, of this relation to the data in Fig.~\ref{fig:mass_sig}.  We perform the fit using the MPFIT routine in IDL \citep{williams+10, markwardt09} and taking into account only the uncertainties in the velocity dispersion (i.e., at fixed \Planck\ SZ mass\footnote{Taking into account errors on both velocity and mass measurements does not noticeably change the result.}).  

For a robust estimation of the best-fit parameters, we perform 1000 bootstrap resamplings of the pairs ($\Mpl, \sigma_{200}$),  re-computing the best-fit parameters each time.  This yields $A \equiv e^a = (1172 \pm 93)$, and a slope $\alpha= 0.28 \pm 0.20$ (at 68.3\% confidence).  The slope is consistent with the self-similar expectation of $\alpha=1/3$, although with large uncertainty.  We henceforth set $\alpha=1/3$ and refit to find $A=(1158\pm 61)$.  
The dispersion of the velocity measurements about the best-fit line (i.e., at given $\Mpl$) is $\langle \delta^2_{\ln \sigma} \rangle ^{1/2} = 0.189 \pm 0.009$.  The best fit together with the data is plotted in Fig.~\ref{fig:mass_sig}. 
A model with a zero slope is excluded at $\sim2\sigma$ confidence, using the $\chi^2$ difference (the $\chi^2$ for the best-fit model is 12.2, the $\chi^2$ for the zero-slope model is 14.3). 
We also performed the fit using only clusters with greater than 20 member galaxies.  Once again fixing $\alpha=1/3$, we find $A=(1156\pm 58)$, in this case, consistent with the previous value.  

Our estimator for the mass bias then follows from the formalism of Appendix~\ref{sec:appendix} (Eq.~\ref{eq:best}):
\begin{equation}
(1-b) = \left( \frac{A_g}{A} \right)^3 \feb\fr = \left( \frac{A_d}{A} \right)^3 \bvel^3\feb\fr\,,
\end{equation}
where $\feb$ (Eq.~\ref{eq:eb}) is the Eddington bias correction and $\fr$ (Eq.~\ref{eq:rcorr}) is a correction for correlated scatter between velocity dispersion and the \Planck\ mass proxy.  With our value for the normalization fit to the data and the value for dark matter from \citet{evrard+08}, we have numerically,
\begin{equation}\label{eq:mb}
(1-b) = (0.55\pm 0.09) \bvel^3\feb\fr\,. 
\end{equation}
In the next two subsections, we propose $\feb=0.93\pm 0.01$ and $\fr\approx 1.01$ as reference values.
Our final value for the mass bias also depends on the cube of the velocity bias.  Adopting our baseline $\bvel = 1.08$ from \citet{munari+13}, we have
\begin{equation}
\label{eq:munari}
(1-b)   =  (0.64\pm 0.11) \left(\frac{\fr}{1.01}\right) \,.
\end{equation}

The quoted uncertainty accounts for measurement error, uncertainty on the Eddington bias correction and uncertainty on the velocity bias given by \citet{munari+13}; it is dominated by the measurement error. The uncertainty on Munari et al.'s velocity bias ($\sim 0.6\%$) is a negligible contribution to our total error budget.
  It is more difficult to assign an uncertainty to the correction for correlated scatter, as this depends on the details of cluster physics; we argue below that feedback makes this a minor correction, as reflected in our fiducial value of $\fr = 1.01$.   \\
A summary of best-fit parameters is provided in Table \ref{summary} for several velocity dispersion--mass relations. Where the slope is set to 1/3, we quote our estimates of the \Planck\ mass bias for the velocity bias derived by \citet{munari+13}, $\bvel=1.08$. We distinguish results for the full sample from results for the subsample of clusters with at least 20 member galaxies.

Our value of $(1-b)   =  0.64\pm 0.11$ lies within $1\sigma$ of the value $(1-b) = 0.58\pm 0.04$ needed to  reconcile the cluster counts with the primary CMB constraints.

\begin{table*}
\begin{center}
\caption{Best-fit values and vertical scatter (i.e., at given mass) of the velocity dispersion--mass relation, $\sigma=A[E(z)M/10^{15}M_\odot]^B$, together with mass bias estimates. 
Results are given for our velocity dispersion estimates, $\sigv(<R_{\text{max}})$, and for the derived velocity dispersions within $R_{200}$, $\sigma_{200}$. 
We distinguish the case where all clusters in the sample are included in the fit from the case where only those with at least 20 member galaxies are considered. \label{summary}}
\vspace{0.25cm}
\resizebox{!}{3.6cm}{
\begin{tabular}{c c c c c c}
\tableline \\
Relation & A & B & scatter & $(1-b)/\bvel^3\feb\fr\ $&$(1-b)^\text{a}_\text{Munari}$ \\
&(km~$\rm s^{-1}$)&&$\langle \delta^2_{\ln \sigma} \rangle ^{1/2}$&&\\
\tableline \\
\it All clusters \\
\tableline \\
$\sigv(<R_{\text{max}})-M^\text{Pl}_{200}$&$1239\pm99$&$0.29\pm0.21$&$0.189\pm0.018$&--&--\\
$\sigv(<R_{\text{max}})-M^\text{Pl}_{200}$&$1226\pm68$&$1/3$&$0.182\pm0.012$&$0.47\pm0.08$&$0.55\pm0.09$\\
$\sigma_{200}-M^\text{Pl}_{200}$&$1172\pm93$&$0.28\pm0.20$&$0.198\pm0.018$&--&--\\
$\sigma_{200}-M^\text{Pl}_{200}$&$1158\pm61$&$1/3$&$0.189\pm0.009$&$0.55\pm0.09$&$0.64\pm0.11$\\
\tableline \\
\it Only clusters with $N_\text{gal}\ge20$\\
\tableline \\
$\sigv(<R_{\text{max}})-M^\text{Pl}_{200}$&$1250\pm71$&$1/3$&$0.168\pm0.014$&$0.44\pm0.08$&$0.51\pm0.09$ \\
$\sigma_{200}-M^\text{Pl}_{200}$&$1156\pm58$&$1/3$&$0.136\pm0.012$&$0.56\pm0.08$&$0.66\pm0.09$\\
\end{tabular}
}
\end{center}
\footnotetext{The values of the mass bias quoted in the last column are obtained using the velocity bias, $\bvel\ $, derived by \cite{munari+13}, following the notation of Eq. (\ref{eq:munari}), where the Eddington bias correction is also included.}
\end{table*}

\subsection{Eddington Bias}
\label{sec:eb}
In this section, we detail our Eddington bias correction. The Eddington bias correction (Eq.~\ref{eq:eb}),
\begin{equation}
\feb = e^{-\beta\TdisMpl^2},
\end{equation}
depends on the local slope of the mass function on cluster scales, $\beta\approx 3$, and the total dispersion, $\TdisMpl$, of the \Planck\ mass proxy at fixed true mass.  This is because we assume that our sample is a random draw from the parent sample selected on $\Mpl$.  As described in Sec.~\ref{sec:Mpl}, the mass proxy is calculated as an intersection of \Planck\ SZ measurements and the X-ray based scaling relation in \citet{PlanckSZCosmo2014}.  We characterize the measurement uncertainty on $\Mpl$ by averaging the calculated uncertainty over our cluster sample: $\MdisMpl=0.13\pm 0.02$.  To estimate the intrinsic scatter, we convert the $0.17\pm 0.02$ dispersion of the $Y-M^{5/3}$ relation \citep{PlanckSZCosmo2014} to $\disMpl=(3/5)(0.17\pm 0.02)=0.10\pm 0.01$.  Combining the two, we arrive at a total scatter of 
\begin{equation}
\label{eq:ms}
\TdisMpl=0.16\pm 0.02 \,.
\end{equation}
 
Setting $\beta=3$, we calculate an Eddington bias correction of
\begin{equation}
\ln\feb = -0.08(1\pm 0.19),
\end{equation}
or a reference value of $\feb=0.93(1\pm 0.01) = 0.93\pm 0.01$.  

Our estimate for the intrinsic scatter in the \Planck\ mass from \citet{PlanckSZCosmo2014} may be optimistic.  If we allow a value 50\% larger, we get a correction of $\feb=0.84\pm0.027$. The resulting mass bias would be $(1-b)=(0.58\pm0.097) (\fr/1.01)$.

\subsection{Correlated Scatter}
\label{sec:corr}
The second correction to our mass bias estimator arises from correlated scatter between velocity dispersion and the \Planck\ mass proxy.  It is given by (Eq.~\ref{eq:rcorr}),
\begin{equation}
\fr     =  e^{3\corr\beta\disv\disMpl},
\end{equation}
because only the intrinsic scatter is correlated.  \citet{stanek+10} examined the covariance between different cluster observables using the Millennium Gas Simulations \citep{hartley+08}.  They found significant intrinsic correlation between velocity dispersion and SZ signal, $\corr = 0.54$, in the simulation with only gravitational heating.  In the simulation additionally including cooling and pre-heating, however, the correlation dropped to $\corr=0.079$.  This would seem to make sense as we might expect non-gravitational physics, such as feedback and cooling, to decouple the SZ signal, which measures the total thermal energy of the gas, from the collisionless component.  

While the scatter of the dark matter velocity dispersion is only 4\%, \citet{munari+13} find a scatter in the range $0.1-0.15$ for their subhalos and galaxies. 
Fixing $\beta=3$ and taking $\corr=0.08$, $\disv=0.15$ and $\disMpl=(3/5)0.17=0.10$ as reference values, we have
\begin{equation}
\ln\fr = 0.010\left(\frac{\corr}{0.08}\right)\left(\frac{\disv}{0.15}\right)\left(\frac{\disMpl}{0.10}\right),
\end{equation} 
or a reference value of $\fr=1.01$.  

\section{Discussion}
\label{sec:discussion}
We have estimated the \Planck\ cluster mass bias parameter by measuring the velocity dispersion of 17 SZ-selected clusters observed with Gemini.  It is corrected for both Eddington bias and possible correlated scatter between velocity dispersion and the SZ mass proxy.  These corrections are based on a multivariate log-normal model for the cluster observables that is detailed in Appendix~\ref{sec:appendix}.  We do not correct individual cluster masses for Eddington bias \citep[e.g.,][]{sifon+16}, but rather apply a global correction to the mean scaling relation between velocity dispersion and \Planck\ mass proxy.   

Our primary objective in calibrating the mass bias of \Planck\ clusters is to inform the cosmological interpretation of the \Planck\ cluster counts. \citet{PlanckSZCosmo2014} and \citet{planck_xxiv} found tension between the observed cluster counts and the counts predicted by the base $\Lambda$CDM model fit to the primary CMB anisotropies, with the counts preferring lower values of the power spectrum normalization, $\sigma_8$.  The importance of the tension, however, depends on the normalization of the SZ signal -- mass scaling relation.  The \Planck\ team uses a relation calibrated on \XMM\  observations of clusters \citep[see the Appendix of][]{PlanckSZCosmo2014}, and proposed the mass bias parameter, $b$, to account for possible systematic offsets in this calibration due to astrophysics and (X-ray) instrument calibration.  No offset corresponds to $b=0$, while the value needed to reconcile the observed cluster counts with the base $\Lambda$CDM model is $(1-b)=0.58\pm0.04$ \citep{planck_xxiv}.

The possible tension between clusters and primary CMB has motivated a number of recent studies of the cluster mass bias in both X-ray and SZ catalogues \citep[e.g.,][]{sifon+13,sifon+16,ruel+14,bocquet+15, battaglia+15,simet+15,smith+16}.  For a like-to-like comparison, we focus here on determinations for the \Planck\ clusters.  

\citet{rines+16} compare SZ and dynamical mass estimates of 123 clusters from the \Planck\ SZ catalog in the redshift range $0.05<z<0.3$. They use optical spectroscopy from the Hectospec Cluster Survey \citep{rines+13} and the Cluster Infall Regions in SDSS project \citep{rines_diaferio2006}, observing a velocity dispersion--SZ mass relation in good agreement with the virial scaling relation of dark matter particles. They find neither significant bias of the SZ masses compared to the dynamical masses, nor evidence of large galaxy velocity bias.  They conclude that mass calibration of \Planck\ clusters can not solve the CMB--SZ tension and another explanation, such as massive neutrinos, is required.  

\citet{vdl+14} examine 22 clusters from the Weighing the Giants (WtG) project that are also used in the \Planck\ cluster count cosmology analysis. Applying a weak lensing analysis, they derive considerably larger masses than \Planck, measuring an average mass ratio of $\langle M_{\rm Planck}/M_{\rm WtG}\rangle = 0.688\pm0.072$ with decreasing values for larger \Planck\ masses. They claim a mass-dependent calibration problem, possibly due to the fact that the X-ray hydrostatic measurements used to calibrate the \Planck\ cluster masses rely on a temperature-dependent calibration.  A similar result is obtained by \citet{hoekstra+15} based on a weak lensing analysis of 50 clusters from the Canadian Cluster Comparison Project (CCCP). For the clusters detected by \Planck, they  find a bias of $0.76 \pm 0.05\text{(stat)} \pm 0.06\text{(syst)}$, with the uncertainty in the determination of photometric redshifts being the largest source of systematic error.  \citet{planck_xxiv} used these latter two measurements as priors in their analysis of the SZ cluster counts.  They also employed a novel technique based on CMB lensing \citep{mb15} to find $1/(1-b)=0.99\pm 0.19$ when averaged over the full cluster cosmology sample of more than 400 clusters.  As later pointed out by \citet{battaglia+15}, these constraints should be corrected for Eddington bias\footnote{There is some confusion in the nature of these corrections. \citet{battaglia+15} propose a correction for WtG and CCCP that is really more akin to a Malmquist bias, i.e., due to selection effects arising from the fact that  some clusters in the WtG and CCCP samples do not have \Planck\ mass proxy measurements.}.

\citet{smith+16} use three sets of independent mass measurements to study the departures from hydrostatic equilibrium in the Local Cluster Substructure Survey (LoCuSS) sample of 50 clusters at $0.15 < z < 0.3$. The mass measurements comprise weak-lensing masses \citep{okabe+smith2016, ziparo+15}, direct measurements of hydrostatic masses using X-ray observations \citep{martino+14}, and estimated hydrostatic masses from \citet{Planck2015}. They found agreement between the X-ray-based and \Planck-based tests of hydrostatic equilibrium, with an X-ray bias of $0.95 \pm 0.05$ and an SZ bias of $0.95 \pm 0.04$.  

Finally, \citet{penna-lima+2016} used lensing mass measurements from the Cluster Lensing And Supernova  \citep[CLASH,][]{postman+12}  survey with Hubble  to find a \Planck\ mass bias of $(1-b)=0.73\pm 0.10$.  Employing a Bayesian analysis, they modeled the CLASH selection function and astrophysical effects, such as scatter in lensing and SZ masses and their potential correlated scatter, as well as possible bias in the lensing measurements.  Their quoted uncertainty accounts for these effects by marginalizing over the associated nuisance parameters. They also provide a summary of recent mass calibration measurements, including the Eddington bias correction proposed by \citet{battaglia+15} for the WtG and CCCP determinations.
\citet{sereno+17}  found a result similar to Penna--Lima for the \Planck\ mass bias $(1-b) = 0.76\pm0.08$, using weak lensing masses from the Canada France Hawaii Telescope Lensing Survey \citep[CFHTLenS,][]{heymans+12} and the Red Cluster Sequence Lensing Survey \citep[RCSLenS,][]{hildebrandt+16}.

Comparing to the values above, our results  is $\sim 30\%$ lower (at $\sim 2.5 \sigma$) than both the \citet{smith+16} lensing determination and the
\citet{rines+16} determination, also based on velocity dispersions, both of which favor little or no mass bias.   However, we agree within 1$\sigma$ with the results from WtG \citep{vdl+14}, the CCCP \citep{hoekstra+15} and the CLASH \citep{postman+12} analysis by \citet{penna-lima+2016}. 

 If we use our value of $(1-b)=(0.58\pm0.097) (\fr/1.01)$, obtained with 50\% larger intrinsic scatter on \Planck\ masses (see Sect.~\ref{sec:eb}), it would still agree within 2$\sigma$ with the results from weak lensing cited above. In both cases, our value of the mass bias is within 1$\sigma$ of the value $(1-b) = (0.58\pm 0.04)$ needed to reconcile the cluster counts with the primary CMB.

\subsection{Estimating the velocity bias $\bvel$ using a prior on the mass bias}
\label{sec:bv_calc}
Given the large differences in the velocity bias as predicted by simulations, it is worth turning the vice -- the strong dependence of our mass calibration on velocity bias -- into a virtue:  relying on accurate mass estimates provided by weak lensing analyses, we derive a constraint on $\bvel$ from our measured velocity dispersions.  We adopt the \Planck\ mass calibration obtained by \citet{penna-lima+2016}, based lensing mass measurements from the Cluster Lensing And Supernova survey with Hubble (CLASH).  Using a Bayesian analysis of CLASH mass measurements and \Planck\ SZ measurements, they marginalize over nuisance parameters describing the cluster scaling relations and the sample selection function to obtain $(1-b)=0.73\pm 0.10$. This is a reasonable prior, since the  \citet{penna-lima+2016} sample is characteristic in mass (and we also assume in mass bias) of \Planck\ detected clusters. Using this as a prior on the mass bias in Eq. (\ref{eq:mb}), with our reference value for the Eddington bias given in Section \ref{sec:eb}, we then deduce the constraint 
\begin{equation}
\bvel=1.12\pm0.07  \left(\frac{1.01}{\fr}\right)^{1/3} \,.
\end{equation}
This positive velocity bias agrees with the value from the \citet{munari+13} simulations and the \citet{guo+15} result for samples more luminous than $M_\text{r}=20.5$ ($L_\star$).  It is reasonably consistent (within $2\sigma$) with the results of \citet{wu+13} that predict nearly unbiased velocities for the brightest 10-30 galaxies, appropriate for our sample.  Our result is discrepant, at $3\sigma$, with the negative velocity bias $\bvel \lesssim 0.9$, as for example found by \cite{caldwell+16} simulations.

\section{Conclusions}
\label{sec:conclusions} 
We have examined the \Planck\ cluster mass bias using a sample of 17 \Planck\ clusters for which we measured velocity dispersions with GMOS at the Gemini observatory.  The unknown velocity bias, $\bvel$, of the member galaxy population is the largest source of uncertainty in our final result:  $(1-b)=(0.51\pm0.09) \bvel^3$.
Using our baseline value for $\bvel$ from \citet{munari+13}, we find $(1-b)=(0.64\pm 0.11)$, consistent within just over 1$\sigma$ with WtG, CCCP and CLASH, and within 1$\sigma$ of the value $(1-b) = (0.58\pm 0.04)$  needed to reconcile the \Planck\ cluster counts with the primary CMB.

We conclude that velocity bias is the primary factor limiting interpretation of dynamical cluster mass measurements at this time.  It is essential to eliminate this modeling uncertainty if velocity dispersion is to be a robust mass determination method.  

Turning the analysis around, observational constraints on the velocity bias can be obtained by combining accurate mass estimates from weak lensing measurements with velocity dispersion measurements.  Assuming a prior on the mass bias from \citet{penna-lima+2016}, we derive $\bvel = 1.12 \pm 0.07$, consistent with our baseline value from \citet{munari+13} ($\bvel = 1.08$) and with results from \citet{wu+13} and  \citet{guo+15}, but discrepant at $3\sigma$ with negative velocity bias $\bvel \lesssim 0.9$, as for example found by \cite{caldwell+16}.

Apart from modeling uncertainty on the velocity bias, we have achieved a precision of 17\% on the mass bias measurement with 17 clusters.  Assuming that the simulations will eventually settle on a value for the velocity bias, this motivates continued effort to increase our sample size to produce a 10\% or better determination, comparable to recent weak lensing measurements. 

\acknowledgments{We thank our referee, Gus Evrard, for constructive discussion that helped improve the presentation of this work. We thank Andrea Biviano and Ian McCarthy for useful discussions. Based on observations obtained at the Gemini Observatory (Programs GN-2011A-Q-119, GN-2011B-Q-41, and GS-2012A-Q-77; P.I. J.G. Bartlett), which is operated by the Association of Universities for Research in Astronomy, Inc., under a cooperative agreement with the NSF on behalf of the Gemini partnership: the National Science Foundation (United States), the National Research Council (Canada), CONICYT (Chile), Ministerio de Ciencia, Tecnolog'a e Innovaci—n Productiva (Argentina), and MinistŽrio da Cincia, Tecnologia e Inova‹o (Brazil). Supported by the Gemini Observatory, which is operated by the Association of Universities for Research in Astronomy, Inc., on behalf of the international Gemini partnership of Argentina, Brazil, Canada, Chile, and the United States of America. This material is based upon work supported by AURA through the National Science Foundation under AURA Cooperative Agreement AST 0132798 as amended. J.G.B. and S.M. acknowledge financial support from the {\em Institut Universitaire de France (IUF)} as senior members. 
The work of J.G.B., C.L. and D.S. was carried out at the Jet Propulsion Laboratory, California Institute of Technology, under a contract with NASA. S.M.'s research was supported by an appointment to the NASA Postdoctoral Program at the Jet Propulsion Laboratory, administered by Universities Space Research Association under contract with NASA.

\facility{Gemini: South, Gemini: Gillet, Hale, Planck}
\newpage
\bibliographystyle{aasjournal}
\bibliography{mybib}

\appendix
\section{Conversion from $\mMpl\ $ to $\Mpl\ $}
\label{sec:conversion}
To compare our mass measurements to other independent estimates, we rescale the \Planck\ masses to $\Mpl\ $ using the mass-concentration relation of \citet{dm14}.  This relation is derived from N-body simulations of relaxed dark matter halos in a \Planck\ cosmology, as adopted here. It is in good agreement with the recently proposed universal model of \citet{DK2015}, which includes both relaxed and unrelaxed halos, for the mass and redshift range of interest. 

We assume a Navarro-Frenk-White \citep[NFW,][]{nfw97} density profile, and we choose an input value for the concentration $c_{200} = 5$, which is consistent with the model of \citet{dm14} for a $10^{15}h^{-1}\,M_\odot$ cluster in the redshift range $0<z<0.5$.  We then convert to $\Mpl\ $:
\begin{equation}
\Mpl\  = \mMpl\ \frac{f(c_{200})}{f(c_{500})} \,,
\label{eq:ini}
\end{equation}
where $f(c_\Delta)=\log(1+c_\Delta) - \frac{c_\Delta}{1+c_\Delta}$ indicates a general density contrast.
We calculate $c_{500}$ from
\begin{equation}
\mMpl\ = 4\pi \rho_s r_s^3 f(c_{500}),
\label{eq:m500}
\end{equation}
where $c_{500}$ is the only unknown quantity, because the scale density parameter, $\rho_s$, is fixed by the NFW profile,
\begin{equation}
\rho_s = \rho_{c,z} \frac{200}{3} \frac{c_{200}^3}{\ln(1+c_{200})-\frac{c_{200}}{1+c_{200}}}, 
\end{equation}
and the scale radius is
\begin{equation}
r_s = \frac{R_{500}}{c_{500}} \,,
\end{equation}
with
\begin{equation}
R_{500} = \left[ \mMpl\ \frac{3}{4\pi} \frac{1}{500\, \rho_{c,z}} \right]^{1/3} \,.
\label{eq:fin}
\end{equation}

We solve Eq.~(\ref{eq:m500}) for $c_{500}$ using the ZBRENT.PRO routine in IDL and obtain a first estimate of $\Mpl\ $ from Eq. (\ref{eq:ini}).  We then use the mass-concentration relation in Eq.~(8) of \citet{dm14} to get a new value for $c_{200}$.  We iterate this algorithm until we reach 5\% accuracy on $\Mpl\ $  (i.e., the difference between the mass estimated at the iteration {\it i} and the mass estimated at the iteration {\it i}-1 is less than 0.05).  We find smaller concentrations than the starting value of 5, with a mean $c_{200}=4.2$.  We have verified that the algorithm converges to the same values of $\Mpl$ when changing the initial input value of $c_{200}$.  

We implemented this procedure in a Monte Carlo simulation with 1000 inputs for each cluster, sampling the \Planck\ mass, $\mMpl\ $, according to a normal distribution with a standard deviation taken as the geometric mean of the uncertainties listed in Table~\ref{results}.  Similarly, we consider a log-normal distribution for $c_{200}$ with a mean given by Eq.~(8) in \citet{dm14} and standard deviation equal to the intrinsic scatter of 0.11 dex in the mass--concentration relation.  This yields a log-normal distribution of calculated $\Mpl\ $ values from Eq.~(\ref{eq:ini}), whose mean and standard deviation are also listed in Table~\ref{results}.

\section{Cluster Model}
\label{sec:appendix}
To construct an estimator for the mass bias, we adopt a multivariate log-normal model for the cluster observables $\sigv$ and $\Mpl$ at fixed true mass, $\Mt$, following \citet{white+10,stanek+10} \citep[see also,][]{allen+11,rozo+14b,evrard+14}.  It is then convenient to work with the logarithm of these quantities: $\lsigv = \ln(\sigv/\text{km\,s}^{-1})$, $\lMpl = \ln(E(z)\Mpl/10^{15}\,\Msol)$ and $\mu = \ln(E(z)\Mt/10^{15}\,\Msol)$, where we incorporate self-similar evolution with redshift, $E(z)$, with the masses.
Power-law scaling relations give the observable mean values at true mass as,
\begin{eqnarray}
\mlMpl  \equiv \langle \lMpl | \mu \rangle & = & \ln(1-b) + \mu,\\
\mlsigv \equiv \langle \lsigv | \mu \rangle & = & \av + \alphav\mu,
\end{eqnarray}
where the averages are taken over both intrinsic cluster properties and measurement errors.  The first relation is simply our definition of the mass bias, Eq.~(\ref{eq:massbias}), and in practice we take $\alphav=1/3$, its self-similar value, in the second relation.

Each observable is also associated a log-normal dispersion about its mean that includes both intrinsic and measurement scatter:
\begin{eqnarray}\label{eq:disp}
\Tdisv^2     & = & \disv^2 + \Mdisv^2,\\
\TdisMpl^2 & = & \disMpl^2 + \MdisMpl^2,
\end{eqnarray}
where the first terms are the intrinsic log-normal scatter and the second ones are the measurement error.  Although measurement error is Gaussian in the observed quantity, rather than log-normal, we treat its fractional value as a log-normal dispersion; this is an approximation good to first order in the fractional measurement error.  The second terms in the above expressions will therefore be understood as fractional measurement errors.  The intrinsic dispersions may be correlated with correlation coefficient $\corr=\langle (\lsigv-\mlsigv)(\lMpl-\mlMpl) \rangle/(\disv\disMpl)$.

It is then possible to show that the predicted scaling between velocity dispersion and \Planck\ mass is
\begin{equation}
\langle \lsigv | \lMpl \rangle = \av + \alphav\left[\lMpl-\ln(1-b)  - \beta \TdisMpl^2 + r\beta\alphav^{-1}\Tdisv\TdisMpl\right],
\end{equation}
where $\beta$ is the slope of the mass function on cluster scales, $\beta\approx 3$.  The second to last term is the Eddington bias, proportional to the full dispersion, intrinsic and measurement, in the sample selection observable, $\lMpl$.  In the last term, $r=\corr(\disv/\Tdisv)(\disMpl/\TdisMpl)$, i.e., the intrinsic correlation coefficient diluted by the measurement errors.  The last term is therefore equivalent to $\corr\beta\alphav^{-1}\disv\disMpl$.

This is the prediction for our measured scaling relation.  Comparison to our fit identifies
\begin{eqnarray}
\ln A & = & {\av - \alphav\left[\ln(1-b)+\beta\TdisMpl^2 - \corr\beta\alphav^{-1}\disv\disMpl \right]},
\end{eqnarray}
which leads to our estimator
\begin{equation}\label{eq:best}
(1-b) = \left(\frac{\Ag}{A}\right)^3 \feb \fr,
\end{equation}
with 
\begin{eqnarray}
\label{eq:eb}
\feb & = & e^{-\beta\TdisMpl^2},\\
\label{eq:rcorr}
\fr    & = & e^{3\corr\beta\disv\disMpl}, 
\end{eqnarray}
after setting $\alphav=1/3$.  As expected, the Eddington bias correction increases true cluster mass at given $\Mpl$, increasing the mass bias, $b$ (decreasing $1-b$).  A positive correlation between velocity dispersion and \Planck\ mass has the opposite effect.


\listofchanges

\end{document}